\documentstyle[12pt,epsf]{article}

\newcommand{\beq}{\begin{equation}}
\newcommand{\eeq}{\end{equation}}
\newcommand{\bea}{\begin{eqnarray}}
\newcommand{\eea}{\end{eqnarray}}

\newcommand{\complex}{{\bb C}} %% complex numbers
\newcommand{\zed}{{\bb Z}} %% integers
\newcommand{\real}{{\bb R}} %% real numbers
\newcommand{\ids}{{\bbs I}} %% small identity operator
\newcommand{\tor}{{\bb T}} %% one dimensional torus \real /\zed

\newtheorem{theorem}{Theorem}

\font\mybb=msbm10 at 12pt
\def\bb#1{\hbox{\mybb#1}}
\font\mybbs=msbm10 at 9pt
\def\bbs#1{\hbox{\mybbs#1}}

\newcommand{\newsection}[1]
{\vspace{4mm}
\pagebreak[3]
\addtocounter{section}{1}
\setcounter{equation}{0}
\setcounter{subsection}{0}
\begin{flushleft}
{\large\bf \thesection. #1}
\end{flushleft}
\nopagebreak
\medskip
\nopagebreak}

\newcommand{\newsubsection}[1]{
 \vspace{4mm}
\pagebreak[3]
\addtocounter{subsection}{1}
\noindent{ \bf \thesubsection. #1}
\nopagebreak
\vspace{1.7mm}
\nopagebreak}

\font\mybb=msbm10 at 12pt
\font\mybbs=msbm10 at 9pt

\makeatletter
\newdimen\normalarrayskip              % skip between lines
\newdimen\minarrayskip                 % minimal skip between lines
\normalarrayskip\baselineskip
\minarrayskip\jot
\newif\ifold             \oldtrue            \def\new{\oldfalse}

\makeatother

\newlength{\extraspace}
\newlength{\extraspaces}
\begin{document}

\addtolength{\baselineskip}{.8mm}

\thispagestyle{empty}

\begin{flushright}
\baselineskip=12pt
CBPF-NF-022/99\\
LNCC-RP-013/99\\
hep-th/990????\\
\hfill{  }\\May 1999
\end{flushright}
\vspace{.5cm}

\begin{center}

\baselineskip=24pt

{\Large\bf{ On the Algebraic K-theory of The Massive\\ D$8$ and M$9$ Branes
}}\\[15mm]

\author{\\
Department of Theoretical Physics, State University of Rio de Janeiro\\
Rua Sa\~{o} Francisco Xavier, 524-Maracan\~{a}, Rio de Janeiro-RJ}
 
\date{}

\baselineskip=12pt

{\bf Ion V. Vancea} 
\\[5mm]

{\it
Department of Theoretical Physics, State University of Rio de Janeiro\\
Rua Sa\~{o} Francisco Xavier, 524-Maracan\~{a}, Rio de Janeiro-RJ\\
{\tt vancea@cat.cbpf.br}}
\\[15mm]

\vskip 1 in

{\sc Abstract}

\begin{center}
\begin{minipage}{15cm}

We study the relation between the D$8$-branes wrapped on an orientable
compact manifold $W$ in a massive Type IIA supergravity background and the
M$9$-branes wrapped on a compact manifold $Z$ in a massive d=11 supergravity
background from the K-theoretic point of view. 
By speculationg on the use of the dimensional reduction to relate the two
theories in different dimensions and by interpreting the
D$8$-brane
charges as elements of $K_{0}(C(W))$ and the (inequivalent classes of)
spaces of
gauge fields on the M$9$-branes as the elements of
$K_{0}(C(Z)\times _{{\bar{k}}^{*}}G)$ a connection between charges and gauge
fields is argued to exists. This connection is realized as a map between the
corresponding algebraic K-theory groups.

\end{minipage}
\end{center}

\end{center}

\baselineskip=18pt

\noindent
\vfill
\newpage
\pagestyle{plain}
\setcounter{page}{1}

\newsection{Introduction}

As the result of the analysis of the non-BPS brane states \cite{as} the
picture of the charges of the D$p$-branes wrapped on an (orientable)
compact manifold $W$ as elements of the topological K-theory of $W$
emerged \cite{mm,ew}. The charges of all possible D$p$-brane configurations
actually take value in the abelian groups $K^{0}(Y)$, $K^{-1}(Y)$ and
$KO^{0}(Y)$ for Type IIB, Type IIA and Type I branes, respectively, where $Y$
is the d=10 spacetime and $W \subset Y$. By imposing the tadpole anomaly
cancellation in Type IIB and Type I theories, the groups reduce to
$\widetilde{K^{0}}(Y)$ and $\widetilde{KO^{0}}(Y)$, respectively \cite{ew}.
In Type IIA theory the non-existence of any RR boundary state guarantees that
there is no RR spacetime tadpole anomaly \cite{ph}.

Several important questions have already been addressed in literature in the
frame of this theory. The list includes the possibility of using the
Grothendieck groups and the derived categories in the study of brane charges
\cite{es}, the computation of brane charges in various backgrounds
\cite{hgc,sg}, the analysis of T-duality of non-BPS states \cite{bg,kh,bgh}
and the classification of descent and duality relations among branes
\cite{os}.

Another important problem pointed out in \cite{ew} is the understanding of
the K-theory classification of brane charges from eleven dimensional point
of view. The motivation for suspecting a connection between Type IIA brane
charges and some d=11 objects comes from the remark that, on one hand, the
M-theory compactified on $S^1$ is the Type IIA theory and, on the other hand,
the K-theory group of Type IIA brane charges is $\widetilde{K^{-1}}(Y) =
\widetilde{K^{0}}(Y \times S^1 )$ \footnote{In general one can replace the
$\widetilde{K^0 (Y)}$ group with $K^0 (Y)$ group because the corresponding
K-theories are actually K$_c$-theories, i.e. the vector bundles satisfy, from
physical requirements, some appropriate compact support conditions
\cite{ew}.}. The presence of $S^1$ in both theories suggests that the circle
should be actually the same. However, the major obstruction in realizing this
idea in a concrete manner is the fact that there are no 10-branes in M-theory.
This prevents us from giving a sensible physical interpretation of K-theory
of Type IIA branes in eleven dimensions \cite{ew,ph}. One way to circumvent
this difficulty is to use a K-theory that satisfies the following conditions:
i) it allows a physical interpretation of its elements in d=11 and ii) it
represents the Type IIA D-brane charges in d=10.

A theory that satisfies the two conditions above is the algebraic K-theory
\cite{mk,bb}. The group $K_{0}(C(X))$ classifies the finitely generated
projective $C(X)$-modules which are just the spaces of sections of
vector bundles with base manifold $X$. Since these sections, at their turn,
can be interpreted as gauge fields on $X$, $K_{0}(C(X))$ satisfies i) above.
The condition ii) is automatically satisfied since by construction
$K_{0}(C(W)) = K^{0}(W)$ \cite{bb}.

The other crucial ingredient neccesary to describe the D-brane charges from
the eleven dimensional perspective is a map between the branes and the
corresponding objects in d=11. This map can alternatively be thought as a map
between Type IIA theory and the d=11 theory in which the objects are defined.
We have seen above that if the algebraic K-theory is to be used the objects
are the spaces of gauge fields. Then, as the previous discussion and the
original formulation of the problem suggests, the sought for theory in d=11
should be M-theory or a related one. Another way to think of this is to notice
that the map between d=10 and d=11 theories should appear from a natural
connection between these. For M-theory and Type IIA theory
there is such of connection given by the dimensional reduction.

In the following we shall illustrate these ideas on a system formed from
D$8$ and M$9$ branes wrapped on compact manifolds which are embedded in 
massive Type IIA and massive d=11 supergravity backgrounds, respectively. The
massive d=11 supergravity is the one proposed in \cite{blo}. The motivation
for choosing this system relies on the following known facts. The D$8$-brane
is the highest stable Type IIA D-brane and its charges take value in
$K^0 (W)$. Even if 8-brane-antibrane configurations do not contain all lower
dimensional brane configurations, as was pointed out in \cite{ph}, they
contain lower dimensional brane charges. This makes $K^0 (W)$ a nontrivial
interesting object. It is also known that a D$8$-brane can be obtained from a
M${9}$-brane by double dimensional reduction \footnote{The M${9}$ branes
have been initially discussed in \cite{brgpt, hs,hsw,hw,pt,jp,mjd}.}. A
M$9$-brane moves freely in a massive d=11 supergravity background with a
Killing isometry \cite{ch,blo,bs}. The massive d=11 supergravity is connected
to the massive Type IIA supergravity by dimensional reduction. Moreover, its
solitonic solutions include all M-branes from which all Type IIA branes can
be obtained by direct or double dimensional reductions. \footnote{For earlier
discussions on massive Type IIA supergravity see \cite{ljr,hlw,pw}.}
We analyse the possibility of using the dimensional reduction to
connect the objects of interest in the two theories.
In order
to construct a map between the spaces of gauge fields on the M$9$-brane and
the D$8$-brane charges 
have to associate to the dimensional reduction a
geometrical map. In the most favorable case the same map acts simultaneously
between the backgrounds and the spaces on which the branes are
wrapped. As a general case we will assume that the dimensional reduction map
acts only between the backgrounds.  We note that because of the presence of
the Killing isometry, the most natural sections of vector bundles on d=11
background and its compact submanifold are the covariant ones. These are
(free generated) projective modules over the appropriate crossed algebras
\cite{bb}.

The organization of the paper is as follows. In Section 2 we resume some of
the properties of massive Type IIA supergravity and massive d=11 supergravity.
The map between the manifolds and the algebras and the apropriate algebraic
K-theory groups are given in Section 3. In Section 4 we determine the
correspondence between the K-theory groups of gauge fields on the M$9$-brane
and the D$8$-brane charges for various types of Killing vectors. The last
section is devoted to discussions. Two important mathematical results which
are used throughtout the paper are collected in an Appendix.

While this paper has been finished we learned about an interesting analysis
of massive Type II configurations in \cite{jk}. 

\newpage

\newsection{D$8$ and M$8$ branes in massive supergravity backgrounds}

The massive Type IIA supergravity has the following bosonic field content: the
metric $g$, the dilaton field $\phi$, the RR one-form field $C^{(1)}$ which
plays the role of a Stueckelberg field, a massive two-form field $B$ and a
three-form potential $C^{(3)}$. The massless Type IIA supergravity is
obtained from the above theory in two steps. Firstly, one has to redefine the
fields in order to remove the $m^{-1}$ terms present in the supersymmetry
transformation. Secondly, one has to take the limit $m=0$. The theory displays,
in string metric, a cosmological constant that does not depend on the dilaton.
This suggests that the mass parameter can be viewed as the expectation value
of the dual of a RR ten-form field strength.

There is a natural RR nine-form field $A_9$ in the spectrum of the Type IIA
string theory. In order to formulate the massive Type IIA supergravity with
this form field the mass parameter $m$ should be replaced with a mass scalar
field $M(x)$ that obeys the constraint $dM(x)=0$. The field $A_9$ does not
introduce new degrees of freedom and enters a Lagrangian term of the form
$ \sim dA_9 M$. Thus, the field strength of $A_9$ is a Lagrange multiplier for
the constraints of $M(x)$. The solitonic solutions of the massive Type IIA
supergravity with the nine-form field include the ones of massless as well as
of massive Type IIA supergravity. In particular, the solution that carries the
$A_9$ charge is the D$8$-brane. Because the solutions of this theory includes
the Type IIA branes in a concise and elegant manner,
it was proposed that it should be considered as the effective field theory
of the Type IIA superstring \cite{brgpt}.
For latter reference we give here the bosonic part of the massive Type IIA
supergravity, in string frame, which has
the following form \cite{ljr,brgpt}
\begin{eqnarray}
S[g,\phi ,C^{(1)}, B, C^{(3)}]
&=& \frac{1}{16 \pi G^{10}_{N}} \int d^{10}x \sqrt{|g|}
\{ e^{-2\phi}[R(\omega ) -4 (\partial \phi )^2 + \frac{1}{2\dot 3!}H^2 ]
\nonumber\\
&-&[\frac{1}{4}(G^{(2)})^2 + \frac{1}{2 \dot 4!}(G^{(4)})^2 + \frac{1}{8}]
\nonumber\\
&+&\frac{1}{144}\frac{\epsilon}{\sqrt{|g|}}[\partial C^{(3)}\partial C^{(3)}B
+\frac{1}{4}m\partial C^{(3)}b^3 + \frac{9}{320}m^2 B^5 ] \},
\label{act2a}
\end{eqnarray}
where the notations are the ones used in \cite{blo}.

The massive d=11 supergravity proposed in \cite{blo} 
has the following bosonic field content: the metric $\hat{g}$ and
the three-form field $\hat{C}$. The theory is characterized by an
apriori given Killing isometry $\hat{k}$ of the background fields, i.e.
${\hat{\pounds}}_{\hat{k}}{\hat{g}}=
{\hat{\pounds}}_{\hat{k}}{\hat{C}}=0$. It is important to note that,
as a consequence, there is a system of coordinates in which
$\hat{k}^{{\hat{\mu}}} = \delta^{{\hat{\mu}}y}$ and the fields do not depend
on the coordinate $y$ on the integral curve of the Killing vector. In order
to obtain the massless d=11 supergravity one has to set the mass parameter
to zero and to restore the $y$ dependence of fields. We recall that the
massless d=11 supergravity represents the effective field theory of M-theory
and the decompactifying limit of massless Type IIA supergravity. At its turn,
the latter is the $g_s \rightarrow \infty $ limit of Type IIA superstring.

The bosonic part of the massive d=11 supergravity action \cite{blo} is given
by
\begin{eqnarray}
\hat{S} [ \hat{g}, \hat{C} ] &=&
\frac{1}{16 \pi G^{11}_{N}} \int d^{11}x \sqrt{|\hat{g}|}
\{ \hat{R} (\hat{\Omega}) - \frac{1}{2\dot 4!} {\hat{G}}^{2} -
\frac{1}{8} m^2 |{\hat{k}}^2|^2 \nonumber\\
&+& \frac{1}{(144)^2 } \frac{\hat{\epsilon}}{\sqrt{|\hat{g}|}}
[ 16 \partial \hat{C} \partial \hat{C} \hat{C} +
9m \partial \hat{C} \hat{C} (i_{\hat{k}} \hat{C})^2 +
\frac{9}{20} m^2 \hat{C}(i_{\hat{k}} \hat{C})^4 ] \},
\label{act11}
\end{eqnarray}
where $\hat{\Omega}$ represents the connection for the massive gauge
transformation.

The M$9$-brane is a solitonic solution of the above massive d=11 supergravity
conjectured to exist for several reasons. One of them is the non-vanishing
time component of the two-form central charge $Z^{\hat{\mu} \hat{\nu}}$ of the
M-theory superalgebra which also points out that the worldvolume field
theory of the M$9$-brane should be N=1 (chiral) supersymmetric \cite{ch,pkt}.
Another argument comes from the analysis of the eleven dimensional
$E_8 \times E_8$ heterotic string \cite{hw}. Finally, the M$9$-brane fills
in the place of the missing d=11 object that should fit into the pattern of
generating D-branes from M-branes \cite{bs,el}
(see also \cite{hs,hsw,pt,jp,mjd}.) The massive d=11 supergravity is a
theory in which the free M$9$-brane moves naturally, i.e. a world-volume
field theory which contains a vector multiplet with a single scalar necessary
to position the brane, can be constructed. This is possible by assuming that
the Killing isometry of the background is an isometry of the world-volume of
the M$9$-brane, too. Consequently, the world-volume field theory of the
M$9$ should be a gauged sigma model. Actually, such of action was recently
proposed by analogy with the MKK-monopole \cite{el}. Like the other M-branes,
the M$9$-brane gives rise to several Type IIA branes by different dimensional
reductions: a D$8$-brane by reduction along Killing isometry and two gauge
sigma models called KK-7A and KK-8A, respectively,
by other reductions \cite{el}.

The Type IIA D$8$-brane can be identified with the solitonic solution of
massive Type IIA supergravity with nine-form which has 9-dimensional
Poincar\'{e} invariance. This was shown in \cite{brgpt} by proving that
the massive 8-brane is T-dual with the Type IIB D$7$-brane as well as
with the D$9$-brane. Another argument is that, in general,
the background between two D$8$-branes is the massive Type IIA supergravity
\cite{jp}.

\newpage

\newsection{Algebraic K-theory of massive branes}

In this section we construct a geometrical dimensional reduction map
between massive Type IIA and d=11 supergravity backgrounds and give
the algebraic K-theory for D$8$-branes and
M$9$-branes and for the backgrounds in which they are
wrapped.

\newsubsection{The Dimensional Reduction Map}

Let us consider a D$8$ brane-antibrane system wrapped on a compact
submanifold $W$. The massive Type IIA supergravity background in which $W$
is embedded is denoted by $(Y;g,\phi ,C^{(1)};B,C^{(3)})$, where $Y$ is
the spacetime manifold on which the equations of motion for the fields of the
massive Type IIA supergravity hold. There is a natural inclusion map
$p:W \rightarrow Y$. In a similar way, we take the M$9$-branes to be wrapped
on a compact submanifold $Z$ in the massive d=11 supergravity background
$(X;\hat{g},\hat{C};\hat{k})$ and the natural inclusion $i:Z \rightarrow X$.

In order to connect the theory in d=11 with the one in d=10, we need a map
between them. Let us consider the possibility of using the dimensional
reduction of the backgrounds. if we denote by $\hat{\Phi}$ the fields
$\{ \hat{g}, \hat{C} \}$ and by $\Phi$ the fields
$\{ g,\phi,C^{(1)},B,C^{(3)} \}$, the dimensional reduction can be thought
as the a map
\beq
\lambda : \hat{\Phi} \longrightarrow \Phi
\label{dimredmap}
\eeq
which explicitely maps $\lambda (\hat{g}) = (g,\phi , C^{(1)})$ and
$\lambda (\hat{C}) = (B , C^{(3)})$. The precise correspondence between
the components by dimensional reduction is given by \cite{blo}
\begin{eqnarray}
{\hat{g}}_{yy}(\hat{x} ) &=& - e^{\frac{4}{3}\phi (x)}
\nonumber\\
{\hat{g}}_{\mu y}(\hat{x} ) &=& - e^{\frac{4}{3}\phi (x)}C^{(1)}_{\mu }(x)
\nonumber\\
{\hat{g}}_{\mu \nu}(\hat{x} ) &=&  e^{\frac{2}{3}\phi (x)}g_{\mu \nu}(x) -
e^{\frac{4}{3}\phi (x)}C^{(1)}_{\mu }(x)C^{(1)}_{\nu }(x)
\label{drmet}
\end{eqnarray}
for the d=11 metric, and
\bea
\hat{C}_{\mu \nu \rho }(\hat{x}) &=& C_{\mu \nu \rho }^{(3)}(x)
\nonumber\\
\hat{C}_{\mu \nu y }(\hat{x}) &=& B_{\mu \nu }(x)
\label{drform}
\eea
for the d=11 three-form field. If we introduce (\ref{drmet}) and
(\ref{drform}) in the eleven dimensional action (\ref{act11}) then we
obtain the Type IIA action (\ref{act2a}) by using a Palatini type identity.
The two gravitational constants are related by
$G^{10}_{M}=(2\pi l_s )G^{11}_{N}$ since it is assumed that the integral
curve of the Killing vector is a circle of radius equal to the string length
$l_s$ \cite{blo}. Consequently, we can obtain solutions of the equations of
motion
of (\ref{act2a}) from solutions of motion of (\ref{act11}) by dimensional
reduction \cite{bs,el}.
This shows that if we assume the dimensional reduction map (\ref{dimredmap})
between the two background fields in different dimensions
we should extend it to the spacetime manifolds
\beq
\bar{\lambda} : Y \longrightarrow X.
\label{drman}
\eeq
Another way to see that is to notice that $\lambda$ is equivalent to the
explicit relations (\ref{drmet}) and (\ref{drform}). At their turn, these
represent relationships among the component fields, which are functions on
X and Y, respectively. In order to have the equality between
the functions in the both hand sides
of (\ref{drmet}) and (\ref{drform}), a relationship between the
domains of the arguments $Dom(\hat{x})=X$ and $Dom(x)=Y$ must exist. This
relationship is the explicit form of the map $\bar{\lambda}$.

In the adapted coordinate system in which (\ref{drmet}) and (\ref{drform})
are written, the functions on $X$ do not depend on the coordinate $y$.
However,
since $\hat{k}$ is a Killing vector, in principle, one can extend
$\bar{\lambda}$ over the entire manifold $X$ \cite{lpe}. The relation
(\ref{drman}) defines a geometrical map between the two spacetime manifolds.
Depending on the direction on which the dimensional reduction is performed we
can arrive at different manifolds $Y$ from the same $X$. For example, $Y$
can inherit a Killing isometry if the dimensional reduction is performed
along a direction transversal to $\hat{k}$ or does not inherit it if the
direction is along $\hat{k}$.

In a standard fashion, $\lambda$ induces the map between continuous (smooth)
functions
\beq
{\bar{\lambda}}^* : C(X) \longrightarrow C(Y)
\label{funcmap}
\eeq
which is a homomorphism between the C$^*$-algebras of all continuous
functions on $X$ and $Y$, respectively. Actually, like in the case of
$\bar{\lambda}$, the map is not defined over the full $C(Y)$ since the
functions depending on $y$ are left aside. Therefore, (\ref{funcmap})
denotes,
by an abuse of notations, the induced map by the extension over $Y$.

The concrete realisation of the maps (\ref{drman}) and (\ref{funcmap}) 
depends on the choice of the specific solutions of the equations of
motion of the two massive supergravity theories. This choice also
determines their properties. The general analysis that follows relies mainly on the
homomorphism (\ref{funcmap}).

\newsubsection{d=10 Algebraic K-Theory}

The charges of the D$8$-branes belong to the topological K-theory group
$K^{0}(W)$. Its elements are equivalence class of the complex vector bundles
$E\rightarrow W$. Alternatively, they can be interpreted as the elements of
the group $K_{0}(C(W))$ that classifies the spaces of gauge fields over $W$.

In order to see that, let us denote by $\Gamma (E)$ the sections of $E$.
$\Gamma (E)$ can be thought as a finitely generated projective module over
the ring (algebra) $C(W)$ of all complex-valued continuous functions over
$W$. \footnote{Since $K^{0}(W)$ contains information about the topology
of $W$, we work with $C(W)$. If the bundles are real, the module of sections
is defined over $C_{\real}(W)$, the ring of {\em real}-valued functions over
$W$.}. By a theorem by Serre and Swan \cite{ss}, there is a complete
equivalence between the category of vector bundles over a compact space
and the bundle maps and the category of projective modules of finite type
over commutative algebras and module morphisms. In particular, there is an
isomorphism between the monoid of isomorphism classes of complex vector
bundles over $W$ (with Whitney sum) and the monoid of isomorphism
classes of finitely generated projective modules over $C(W)$ ( with ordinary
direct sum) \cite{mk,bb}. This isomorphism suggests that a group similar to
$K^{0}(W)$ could be constructed for the projective modules of sections
of bundles. The construction is standard and goes as follows
\cite{mfa,mk,bb,jlt}.

Let us take the algebra $M_{n}(C(W))$ of $n\times n$ matrices with entries in
$C(W)$. This can be identified with the algebra $C(W,M_n )$ of all continuous
functions from $W$ to the algebra of $n\times n$ complex matrices. Every
idempotent of $C(W,M_n )$ is a finitely generated projective module over
$C(W)$ and it can be obtained from a bundle over $W$. Define next the
inductive limit of finite matrices $M_n (C(W))$ by
\beq
M_{\infty}(C(W)) =  \bigcup_{n=1}^{\infty}M_n (C(W)) \label{indlim}
\eeq
with the natural embedding
\beq
\varphi :M_n (C(W)) \longrightarrow M_{n+1}(C(W))~,
~a \longmapsto \varphi (a) =
\left[ 
\begin{array}{cc}
a & 0 \\ 
0 & 1
\end{array}
\right] .
\label{natemb}
\eeq
Since $C(W)$ is an unital C$^*$-algebra \footnote{The structure of
C$^*$-algebra is given in the supremum norm
$||f(x)||_{\infty} = sup_{x\in W} |f(x)|$. The involution
$*:C(W)\rightarrow C(W)$ is the usual complex conjugation. Unital means
that there is an element $\ids \in C(W)$ such that $f\ids =
\ids f = f
\forall f \in C(W)$. All the algebras are commutative.} every idempotent
$p \in C(W)$ belongs to the set
$ Q(C(W)) = \{ p \in C(W) : \exp (2\pi ip)=1 \}$. Two matrices
$a\in Q_n (C(W))$ and $b\in Q_m (C(W))$ are equivalent if $a$ and $b$ have
similar trivial extensions in some $Q_k (C(W))$
\bea
a \sim b \Longleftrightarrow  & \exists & a'= a \oplus 0_{k-n} \in
Q_k (C(W))~,~\exists b'=b \oplus 0_{k-m} \in Q_{k}(C(W))
\nonumber\\
& \exists &u \in GL_k(C(W)) : a'=ub'u^{-1} ,
\label{equivsim}
\eea
where $GL_n (C(W))$ is the group of all invertible elements of $M_n (C(W))$.
The set of eqivalence classes under $\sim$ in $\cup_{n}Q_n (C(W))$ is an
abelian semigroup denoted by $J(C(W))$. Denote by $UJ(C(W))$ the universal
group of $J(C(W))$ and by $E(C(W))$ its subgroup generated by elements of the
form $ \{ a+b \} - \{ a \} - \{ b \}$, where $\{ a \}$ is the equivalence
class of $a$ in $UJ(C(W))$. Then the first algebraic K-theory group is
defined as
\beq
K_{0}(C(W)) = UJ(C(W))/E(C(W)) .
\label{kzero}
\eeq
The group $K_0 (C(W))$ is the Grothendieck group of the monoid of
isomorphism classes of finitely generated projective $C(W))$ modules.
By construction it satisfies the following identities
\bea
K^0 (W) & = & K_0 (C(W))
\nonumber\\
K^0_{\real} & = & K_0 (C_{\real}(W))
\label{kzeroid}
\eea
for complex and real vector bundles, respectively.

Eq. (\ref{kzeroid}) allows us to interpret the D$8$-brane charges as elements
of either the topological K-theory group $K^0 (W)$ or the algebraic group
$K_0 (C(W))$ and to shift from the complex vector bundles to sections which
have the nice physical interpretation as gauge fields. The amount of
information in both descriptions is the same due to the Serre-Swann theorem.

To $C(W)$ we can associate another algebraic abelian group $K_1 (C(W))$
as follows. We construct firstly the group
\beq
L_n (C(W)) = GL_n (C(W))/GL^0_n (C(W)),
\label{lngroup}
\eeq
where $GL^0_n (C(W)) = \exp (M_n (C(W))$ is the component of identity of
$M_n (C(W))$. The canonical homomorphism
$
GL_n(C(W)) \rightarrow GL_{n+1}(C(W))
$
induced by (\ref{natemb}) yields a homomorphism
$
L_n (C(W)) \rightarrow L_{n+1}(C(W))
$
which makes the sequence $\{ L_n (C(W)) \}$ into a direct limit system
of groups. By definition, $K_1 (C(W))$ is this limit and it is an
abelian group
\beq
K_1 (C(W)) = \lim _{\rightarrow} L_n (C(W)).
\label{kone}
\eeq
We note that, by definiton, $K_1 (C(W))$ takes into account the topology
of $C(W)$\footnote{For any ring $R$ one can construct {\em two} groups
$K_1$ by factorizing $GL_n (R)$ with $GL^0_n (R)$ like in (\ref{lngroup})
or with $E_n (R)$ which is the group of elementary matrices in $GL_n (R)$.
In general, $E_n (R) \in GL^0_n (R)$ and thus the two groups differ. By
construction, only (\ref{lngroup}) contains topological information
about the ring $R$.}.

Since the definitions (\ref{kzero}) and (\ref{kone}) do not depend on a
specific manifold $W$, they can be used for the compact manifold $Z$ and
for compact spacetimes $X$ and $Y$. If $X$ and $Y$ are not compact, the above
definitions no longer apply.

Assume, for definiteness, that $X$ is just a locally compact manifold. The
corresponding algebra $C_0 (X)$ of complex-valued continous functions
vanishing at infinity is non-unital. In this case we take the one point
compactification $X^+$ of $X$ and adjoin the unity to $C_0 (X)$ to form
the algebra $C_0 (X)^+$. The map $C_0 (X) \rightarrow C_0 (X)^+ /C_0 (X)$
determines a complex homomorphism $\varphi_0$ of $C_0 (X)^+$ which yields the
homomorphism
\beq
\varphi^*_0 : K_0 (C(X^+ ))\longrightarrow K_0 (\complex )=0
\label{homzero}
\eeq
The natural algebraic $\widetilde{K}$-theory groups are then defined in
the usual manner by \cite{mk,bb,jlt}
\bea
{\widetilde{K}}_0(C_0 (X)) &=& ker\varphi^*_0
\nonumber\\
{\widetilde{K}}_1(C_0 (X)) &=&K_1 (C_0 (X^+ )).
\label{tildekappa}
\eea

We see that the mathematical construction of the algebraic K-theory groups
is sensible to the topology of the manifolds that characterize our physical
objects. In general, the properties of $K$- and $\widetilde{K}$-theories are
not the same, so we should work with well defined topological spaces. For
simplicity, we will assume in what follows that both spacetimes $X$ and $Y$
are compact. Moreover, when necessary, base points $x_0\in Z \subset X$ and
$y_0 \in W \subset Y$ are understood to be singled out. If the spacetime
manifolds are only locally compact, one should work with their one point
compactifications and with (\ref{tildekappa}) but some extra care should
be taken since, in general, different results are obtained.

\newsubsection{d=11 Algebraic K-Theory}

Even for both $X$ and $Z$ compact, the groups $K_* (C(X))$ and $K_* (C(Z))$ 
are not appropriate to describe the d=11 spacetime and the M9-branes,
respectively. The reason is that there is a Killing isometry $\hat{k}$
in the massive d=11 supergravity background. Also, if we want to obtain the
8-brane from the 9-brane by dimensional reduction, we have to perform it
along the direction of the Killing isometry, which implies that this direction
belongs to the world-volume of the M$9$-brane, too.

The Killing isometry, being a homeomorphism, induces the homeomorphisms
$\hat{k}^*$ and $\bar{k}^*$ of $C(X)$ and $C(Z)$, respectively. Regarding
$\hat{k}$ as the action of a uni-parametric group $G$ on $X$ and $Z$,
$\hat{k}^*$ and $\bar{k}^*$ represent the induced actions on $C(X)$ and
$C(Z)$, respectively. The most appropriate algebras for describing the sets
$(C(X), \hat{k}^* , G)$ and $(C(Z), \bar{k}^* , G)$ are the crossed product
algebras $C(X)\times_{\hat{k}^*}G$ and $C(Z)\times_{\bar{k}^*}G$,
respectively. The crossed product algebra $C(X)\times_{\hat{k}^*}G$ is
defined as the twisted convolution algebra $C_c (G,C(X))$ of continuous
functions from $G$ to $C(X)$ with a natural C$^*$-norm \cite{bb}
\footnote{The involutive algebra is given by
\beq
(\varphi_1 \circ \varphi_2 )(g) = \int \varphi_1 (h) {\hat{k}^*}_h
(\varphi_2 (h^{-1}g))dh
\label{invstr1}
\eeq
for any $\varphi_i \in C_c (G, C(X))$, $g$,$h \in G$, and
\beq
\varphi^* (g)=\delta (g)^{-1}{\hat{k}^*}_g (\varphi(g^{-1})^*),
\label{invstr2}
\eeq
where $\delta : G \rightarrow \real^{*}_+ $ is a modular function on $G$.}.
It is important to note that the unitary representations of the C$^*$-algebra
$C(X)\times_{\hat{k}^*}G$ correspond exactly to the covariant representations
of $(C(X), \hat{k}^*) $. The same remarks are true for $Z$.

In order to make an explicite connection between D$_8$-brane system and the
M$_9$-branes by using the K-theory, it is necessary to establish a
relationship between $C(X)$ and $C(X)\times_{\hat{k}^*}G$. This mathematical
problem has not been solved completely yet for an arbitrary group $G$.
However, important results have been already derived for simply connected
and solvable Lie groups, compact groups and some other groups with a simple
structure \cite{bb}. In what follows we will primarily consider the cases
when $G=\real, \zed $ and $\tor$ where $\tor =\real /\zed$ is the one dimensional
toric group. In the last section we will discuss the more interesting case
when $G$ is an arbitrary compact group.

In the case when $\hat{k}$ represents the action of the real additive group
$G=\real$ on $X$, we use a theorem due to Connes \cite{bb}
that establishes the following isomorphism
\beq
K_p (C(X)) \times_{\hat{k}^*}\real ) \cong K_{p-1}(C(X)),
\label{conthom}
\eeq
where $p=0,1$. Similar isomorphisms exist if $\real $ is replaced by an
arbitrary simply connected and solvable Lie group.

If the group is $G=\zed$, the dual group of the one-dimensional torus,
a theorem by Pimsner and Voiculescu states that the following cyclic
six-term exact sequence connects the algebraic K-theories of $C(X)$ and
$C(X)\times_{\hat{k}^*}\zed$ \cite{bb}
\beq\new{\begin{array}{ccccc}
K_{0}(C(X))&
{\buildrel{1-\hat{k}^*_*}\over\longrightarrow}
&K_{0}(C(X))
&
{\buildrel\iota_*\over\longrightarrow}
&K_{0}(C(X)\times_{\hat{k}^*}\zed )\\& & & &\\
{\scriptstyle\sigma_*}\uparrow& & & &\downarrow
{\scriptstyle\sigma_*}\\& & & &\\K_{1}(C(X)\times_{\hat{k}^*}\zed)
&
{\buildrel\iota_*\over\longleftarrow}
&K_{1}(C(X))
&
{\buildrel{1-\hat{k}^*_*}\over\longleftarrow},
&K_{1}(C(X))\end{array}}
\label{pimvoi}\eeq
where $\hat{k}^*_*$ is the induced map by $\hat{k}^*$ in the K-theory and
$\iota_*$ is the induced map by $C(X)\rightarrow C(X)\times_{\hat{k}^*}\zed$.

A more interesting case is when $G=\tor=\real / \zed$ is the torus group.
Then the following cyclic six-term sequence is exact \cite{bb}
\beq\new{\begin{array}{ccccc}
K_{0}(C(X)\times_{\hat{k}^*}\tor)&
{\buildrel{1-\hat{k}^*_*}\over\longrightarrow}
&K_{0}(C(X)\times_{\hat{k}^*}\tor)
&
{\buildrel{t_*}\over\longrightarrow}
&K_{0}(C(X))\\& & & &\\
{\scriptstyle\rho_*}\uparrow& & & &\downarrow
{\scriptstyle\rho_*}\\& & & &\\K_{1}(C(X))
&
{\buildrel{t_*}\over\longleftarrow}
&K_{1}(C(X)\times_{\hat{k}^*}\tor)
&
{\buildrel{1-\hat{k}^*_*}\over\longleftarrow},
&K_{1}(C(X)\times_{\hat{k}^*}\tor)\end{array}}
\label{pas}\eeq
where $\rho_*$ is induced by the quotient map
\beq
\rho : C(X)\times_{\hat{k}^*}\real \longrightarrow C(X)\times_{\hat{k}^*}\tor.
\label{quot}
\eeq
Here, we wrote explicitely all the maps in the exact sequences for later
reference. Similar sequences exist for $Z$ and $\bar{k}^*$.

Despite the fact that the six-term exact sequences have a limited
general applicability, they provide useful information in many particular
cases. We discuss this problem in the last section.

\newpage

\newsection{The Relationship between Type IIA and d=11 algebraic K-theories}

In this section we are going to derive a map between the groups $K_0(C(W))$
and $K_0 (C(Z)\times_{\bar{k}^*}G)$ which are associated with the
D$8$-brane charges and the spaces of covariant gauge fields on the
M$9$-branes, respectively.

It is instructive to consider first the general situation in which
the dimensional reduction map acts only between $X$ and $Y$. There are
natural inclusion maps for the pair $(X,Z)$
\bea
i &:& Z \longrightarrow X
\nonumber\\
j &:& (X,\emptyset ) \longrightarrow (X,Z)
\label{xzemb}
\eea
and similar ones for the pair $(Y,W)$
\bea
p &:& W \longrightarrow Y
\nonumber\\
q &:& (Y,\emptyset ) \longrightarrow (Y,W).
\label{ywemb}
\eea
A consequence of the Bott periodicity theorem (see Theorem 1 from Appendix)
states that there are the following homomorphisms within the topological
K-theory cohomologies of the pair $(X,Z)$
\bea
r^* &:& K^0 (Z) \longrightarrow K^{-1}(X)
\nonumber\\
l^* &:& K^{-1}(Z) \longrightarrow K^{0}(X),
\label{rlhom}
\eea
given by
\beq
r^* = j^* \circ (\delta \circ \beta ) ~~,~~ l^* = j^* \circ \delta
\label{rlmaps}
\eeq
Here, $\delta$ is the coboundary map and $\beta$ is the Bott isomorphism.
Similar homomorphisms can be constructed for the pair $(Y,W)$, namely
\bea
f^* &:& K^0 (W) \longrightarrow K^{-1}(Y)
\nonumber\\
g^* &:& K^{-1}(W) \longrightarrow K^{0}(Y)
\label{fghom}
\eea
given by the compositions
\beq
f^* = q^* \circ (\omega \circ \tau ) ~~,~~ g^* = q^* \circ \omega,
\label{fgmaps}
\eeq
where the coboundary map is $\omega$ and the isomorphism is $\tau$.

Since the Connes' Thom isomorphism (\ref{conthom}) and the exact sequences
(\ref{pimvoi}) and (\ref{pas}) involve higher groups $K_1$, we need to
connect it with $K^{-1}$. In general, there is no isomorphism between
the two groups \cite{mk,bb}. However, since in our
setup the algebras of continuous functions on $X$, $Z$, $Y$ and $W$ are
unital C$^*$-algebras and since $K_1$ was defined using (\ref{kone})
according to Novodvorskii's theorem, the Gel'fand transform induces an
isomorphism $K_p \rightarrow K^p$, $p=0,1$ for any of these algebras
\cite{men} (see also Theorem 2 from Appendix.) Let us denote these
isomorphisms by $\gamma$ for $X$, $\psi$ for $Z$, $
\varphi$ for Y and
$\epsilon$ for $W$. Then it is easy to verify that we have the following
maps between the algebraic K-theory groups
\bea
m_* &:& K_0(C(Z)) \longrightarrow K_{1}(C(X))~~,~~
n_* : K_1(C(Z)) \longrightarrow K_{0}(C(X))
\nonumber\\
u_* &:& K_0(C(W)) \longrightarrow K_{1}(C(Y))~~,~~
v_* : K_1(C(W)) \longrightarrow K_{0}(C(Y)),
\label{algmaps}
\eea
where $m_*$, $n_*$, $u_*$ and $v_*$ denote the following composition maps
\bea
m_* &=& \gamma \circ r^* \circ \psi^{-1} ~~,
~~n^* = \gamma \circ l^* \circ \psi^{-1}
\nonumber\\
u_* &=& \epsilon \circ f^* \circ \varphi^{-1} ~~,
~~v^* = \epsilon \circ g^* \circ \varphi^{-1}.
\label{explmaps}
\eea
The properties of these maps are esentially the same as of (\ref{rlmaps})
and (\ref{fgmaps}) which are given by the Theorem 1 from Appendix.

While for the Type IIA algebraic groups $u_*$ and $v_*$ represent the end of
the story, in d=11 we have to pass to the algebraic K-theory of the crossed
product algebras as discussed in the precedent section.

Consider firstly the case $G=\real$ and denote by $\chi^X_* $ and
$\chi^Z_* $ the Connes' Thom isomorphisms for $X$ and $Z$, respectively.
By using (\ref{algmaps}) to connect the K-theories of $C(X)$ and $C(Z)$ we
obtain the following composition maps
\bea
\alpha_{* \real} &:& K_{0}(C(Z)\times_{\bar{k}^*}\real)
\longrightarrow K_1 (C(X)\times_{\hat{k}^*} \real)
\nonumber\\
\beta_{* \real} &:& K_{1}(C(Z)\times_{\bar{k}^*} \real)
\longrightarrow K_0 (C(X)\times_{\hat{k}^*} \real),
\label{abrmaps}
\eea
where $\alpha_{* \real}$ and $\beta_{* \real}$ are given by
\bea
\alpha_{* \real} &=& {\chi}^X_* \circ n_* \circ {\chi}^Z_* 
\nonumber\\
\beta_{*\real} &=& {\chi}^X_* \circ m_* \circ {\chi}^Z_*  .
\label{abrmaps1}
\eea
We see that the map between the two sets of abelian groups
$(\alpha_{*\real}, \beta_{*\real})$ is "twisted" in the sense that
it changes the indices 0 and 1.

In the case $G=\zed$, the dual group of $\tor$, we use the Pimsner-Voiculescu
exact sequence (\ref{pimvoi}) for both $X$ and $Z$. Next, as above, we use
$m_*$ and $n_*$ to relate the two sequences. From the sequences obtained
we can read off the following maps
\bea
\alpha_{* \zed} &:& K_{0}(C(Z)\times_{\bar{k}^*}\zed)
\longrightarrow K_0 (C(X)\times_{\hat{k}^*} \zed)
\nonumber\\
\beta_{* \zed} &:& K_{1}(C(Z)\times_{\bar{k}^*} \zed)
\longrightarrow K_1 (C(X)\times_{\hat{k}^*} \zed),
\label{abzmaps}
\eea
which are given by the following compositions
\bea
\alpha_{* \zed} &=& {\iota}_* \circ (1-\hat{k}^*_*) \circ n_*
\circ \sigma_*
\nonumber\\
\beta_{*\zed} &=& {\iota}_* \circ (1- \hat{k}^*_*)\circ m_*
\circ \sigma_*  .
\label{abzmaps1}
\eea
Here, $\iota_* : K_{1}(C(X))\rightarrow K_1 (C(X)\times_{\hat{k}^*} \zed)$
is the map induced by the natural inclusion $\iota : C(X) \rightarrow
C(X)\times _{\hat{k}^* }\zed$ and $\sigma_* : K_1 (C(Z)\times_{\bar{k}^*}
\zed) \rightarrow K_0 (C(Z))$ is the vertical map of the Pimsner-Voiculescu
exact sequence. $\hat{k}^*_*$ denotes the induced map in the algebraic
K-theory by $\hat{k}^*$. Unlike the pair (\ref{abrmaps}) discussed above,
the pair $(\alpha_{*\zed},\beta_{*\zed})$ is not twisted.

We proceed in the similar manner in the case $G=\tor= \real/\zed$. We construct
the cyclic six-term exact sequences for $X$ and $Z$ using (\ref{pas}) and
we connect them with $m_*$ and $n_*$. As a result we obtain the pair
\bea
\alpha_{* \tor} &:& K_{0}(C(Z)\times_{\bar{k}^*}\tor)
\longrightarrow K_0 (C(X)\times_{\hat{k}^*} \tor)
\nonumber\\
\beta_{* \tor} &:& K_{1}(C(Z)\times_{\bar{k}^*} \tor)
\longrightarrow K_1 (C(X)\times_{\hat{k}^*} \tor),
\label{abtmaps}
\eea
where $\alpha_{*\tor}$ and $\beta_{*\tor}$ are composed by
\bea
\alpha_{* \tor} &=& {\rho}_* \circ m_* \circ t_*
\nonumber\\
\beta_{*\tor} &=& {\rho}_* \circ n_* \circ t_*
\label{abtmaps1}
\eea
where $t_* : K_1 (C(Z))\times_{\bar{k}^*}\tor)\rightarrow K_1(C(Z))$ and
$\rho_*$ is the vertical map in the exact sequence (\ref{pas}) of $Z$.
The pair (\ref{abtmaps1}) is not twisted. We also note that $\alpha_{*\tor}$
and $\beta_{*\tor}$ are defined in a minimal way, i.e. without any reference
to the action of $\hat{k}$ on either $X$ or $Z$.

By construction, the properties of $\alpha_*$ and $\beta_*$ are given by
the properties of the component maps for all the groups discussed above. It
is important to notice that, although some of the components of these maps
enter exact sequences for $G=\zed, \tor$ and are Conne's Thom isomorphisms
for $G=\real$, the composition maps do not belong, in general, to exact
sequences due to the presence of $m_*$ and $n_*$. Exacteness is a
supplementary condition which imposes the usual ker/im constraints on the
components.

Let us return to the main problem of determining the map between the relevant
d=10 and d=11 algebraic K-theory groups. We remark that for $G=\zed$ there is
no obvious physical interpretation of the connection between the massive Type
IIA and the massive d=11 supergravity branes and backgrounds in terms of
dimensional reduction map. However, it is instructive to study this case
because the exact sequence (\ref{pas}) in the more interesting case $G=\tor$
is a Pimsner-Voiculescu like sequence. The case $G=\real$ can be thought
of as the decompactifying limit of the Killing direction
$R\rightarrow \infty$.

The starting point is the dimensional reduction map $\bar{\lambda}$ which
induces the homomorphism $\bar{\lambda}^*$ between the unital C$^*$-algebras
$C(X)$ and $C(Y)$. This is assumed to be an unital homomorphism. Due to the
categorical properties of the algebraic K-theory, $\bar{\lambda}^*$ yields
the homomorphisms between the algebraic K-theory groups $K_p(C(X))$ and
$K_p(C(Y))$, $p=0,1$, denoted by $\bar{\lambda}^*_*$. Consequently, it
also connects the exact sequences in d=10 and d=11.

In the case $G=\real$, by connecting the sequences in different dimensions
we obtain the following sequences
\bea
\dots~ 
& \longrightarrow & ~
K_1 (C(Z)\times_{\bar{k}^*}\real)~
\longrightarrow~
K_0 (C(Z))~
\longrightarrow~
\dots~
\longrightarrow~
K_0 (C(X)\times_{\hat{k}^*}\real)~
\nonumber\\
& \longrightarrow & ~
K_1 (C(X))~
\buildrel{\bar{\lambda}^*_*}\over
\longrightarrow~
K_1 (C(Y))~
\longrightarrow~
K_0 (C(W))~
\label{seqrone}
\eea
and
\bea
\dots~ 
& \longrightarrow & ~
K_0 (C(Z)\times_{\bar{k}^*}\real)~
\longrightarrow~
K_1 (C(Z))~
\longrightarrow~
\dots~
\longrightarrow~
K_1 (C(X)\times_{\hat{k}^*}\real)~
\nonumber\\
& \longrightarrow & ~
K_0 (C(X))~
\buildrel{\bar{\lambda}^*_*}\over
\longrightarrow~
K_0 (C(Y))~
\longrightarrow~
K_1 (C(W))~
\label{seqrtwo}
\eea
In general, (\ref{seqrone}) and (\ref{seqrtwo}) are not exact. However, we
can define some composition maps $ a_{*\real}$ and $b_{*\real}$
between the algebraic K-theory groups associated to the wrapped 8-brane
and 9-brane systems
\bea
a_{*\real} &:& K_1 (C(Z)\times_{\bar{k}^*}\real)
\longrightarrow K_0 (C(W))
\nonumber\\
b_{*\real} &:& K_0 (C(Z)\times_{\bar{k}^*}\real)
\longrightarrow K_1 (C(W))
\label{finrmap}
\eea
The expresion of these maps can be easily deduced from the
sequences (\ref{seqrone}) and (\ref{seqrtwo}) and using (\ref{explmaps}) and
(\ref{abrmaps1}) one can see that they have the following form
\bea
a_{*\real}  & = & u^{-1}_* \circ {\bar{\lambda}}^*_*
\circ \chi^X_{*}\circ \beta_{*\real}
\nonumber\\
b_{*\real}  & = & v^{-1}_* \circ {\bar{\lambda}}^*_*
\circ \chi^X_{*}\circ \alpha_{*\real}.
\label{finrmap1}
\eea
The properties of $a_{*\real}$ and $b_{*\real}$ are determined by those of
the components and of $\bar{\lambda}^*_*$. In order to  the sequences
(\ref{seqrone}) and (\ref{seqrtwo}) be exact, $\bar{\lambda}^*_*$ as well
as $u^{-1}_*$, $v^{-1}_*$, $m_*$ and $n_*$ should satisfy the ker/im
exacteness conditions. It is an easy exercise to write them down from
(\ref{explmaps}), (\ref{abrmaps1}) and (\ref{finrmap1}).

In the case $G=\zed$ we can do a similar analysis since $\bar{\lambda}$ has
a geometrical character. However, the discussion in this case is purely formal
and there is no physical interpretation in terms of the dimensional
reduction of the d=11 spacetime $X$ to the d=10 spacetime $Y$. The sequences
built out of $\bar{\lambda}^*_*$ from exact sequences in different dimensions
are given by
\bea
\dots~ 
& \longrightarrow & ~
K_0(C(Z))~
\longrightarrow~
K_0 (C(Z)\times_{\bar{k}^*}\zed)~
\longrightarrow~
K_1 (C(Z))~
\longrightarrow~
\dots~
\nonumber\\
& \longrightarrow & ~
K_0 (C(X)\times_{\hat{k}^*}\zed)~
\longrightarrow~
K_1 (C(X))~
\buildrel{\bar{\lambda}^*_*}\over
\longrightarrow~
K_1 (C(Y))~
\longrightarrow~
K_0 (C(W))~
\nonumber\\
\label{seqzone}
\eea
and
\bea
\dots~ 
& \longrightarrow & ~
K_1 (C(Z))~
\longrightarrow~
K_0 (C(Z)\times_{\bar{k}^*}\zed)~
\longrightarrow~
K_0 (C(Z))~
\longrightarrow~
\dots~
\nonumber\\
& \longrightarrow & ~
K_1 (C(X)\times_{\hat{k}^*}\zed)~
\longrightarrow~
K_0 (C(X))~
\buildrel{\bar{\lambda}^*_*}\over
\longrightarrow~
K_0 (C(Y))~
\longrightarrow~
K_1 (C(W))~
\nonumber\\
\label{seqztwo}
\eea
The composition maps that can be constructed from these sequences act
only geometrically between the K-theories in different dimensions as
follows
\bea
a_{*\zed} &:& K_0 (C(Z)\times_{\bar{k}^*}\zed)
\longrightarrow K_0 (C(W))
\nonumber\\
b_{*\zed} &:& K_1 (C(Z)\times_{\bar{k}^*}\zed)
\longrightarrow K_1 (C(W))
\label{finzmap}
\eea
and they are composed from the following homomorphisms
\bea
a_{*\zed}  & = & u^{-1}_* \circ {\bar{\lambda}}^*_*
\circ \sigma_{*}\circ \alpha_{*\zed}
\nonumber\\
b_{*\zed}  & = & v^{-1}_* \circ {\bar{\lambda}}^*_*
\circ \sigma_{*}\circ \beta_{*\zed}.
\label{finzmap1}
\eea
We emphase once again that presently we cannot claim that (\ref{finzmap1})
have any physical signifiance. However, they are helpful in understanding
the next case.

When $\hat{k}$ represents the action of the group $G=\tor$ on the
manifold $X$, the homomorphism $\bar{\lambda}^*_*$ enters the
following sequences
\bea
\dots~ 
& \longrightarrow & ~
K_0 (C(Z)\times_{\bar{k}^*}\tor)~
\longrightarrow~
K_0 (C(Z))~
\longrightarrow~
\dots~
\nonumber\\
& \longrightarrow & ~
K_1 (C(X)\times_{\hat{k}^*}\tor)~
\longrightarrow~
K_1 (C(X))~
\buildrel{\bar{\lambda}^*_*}\over
\longrightarrow~
K_1 (C(Y))~
\longrightarrow~
K_0 (C(W))~
\nonumber\\
\label{seqtone}
\eea
and
\bea
\dots~ 
& \longrightarrow & ~
K_1 (C(Z)\times_{\bar{k}^*}\tor)~
\longrightarrow~
K_1 (C(Z))~
\longrightarrow~
\dots~
\nonumber\\
& \longrightarrow & ~
K_0 (C(X)\times_{\hat{k}^*}\tor)~
\longrightarrow~
K_0 (C(X))~
\buildrel{\bar{\lambda}^*_*}\over
\longrightarrow~
K_0 (C(Y))~
\longrightarrow~
K_1 (C(W))~
\nonumber\\
\label{seqttwo}
\eea
These sequences define the maps $a_{*\tor}$ and $b_{*\tor}$ between the
corresponding K-groups of the dimensional reduction of the background $X$.
They act as follows
\bea
a_{*\tor} &:& K_0 (C(Z)\times_{\bar{k}^*}\tor)
\longrightarrow K_0 (C(W))
\nonumber\\
b_{*\tor} &:& K_1 (C(Z)\times_{\bar{k}^*}\tor)
\longrightarrow K_1 (C(W))
\label{fintmap}
\eea
From the relations (\ref{explmaps}) and (\ref{abtmaps1}) we obtain the
following expressions
\bea
a_{*\tor}  & = & u^{-1}_* \circ {\bar{\lambda}}^*_*
\circ t_{*}\circ \alpha_{*\tor}
\nonumber\\
b_{*\zed}  & = & v^{-1}_* \circ {\bar{\lambda}}^*_*
\circ t_{*}\circ \beta_{*\tor}.
\label{fintmap1}
\eea
Like in the previous cases, the properties of $a_{*\tor}$ and $b_{*\tor}$
are given by the properties of the components. Another common feature
with the other cases is that the sequences (\ref{seqtone}) and (\ref{seqttwo})
are not, in general, exact.

Note that in the decompactifying limit of the Killing direction,
the 8-brane charges are obtained rather from the $K_1$ associated
to the 9-branes than from $K_0$ whose elements are equivalence classes
of isomorphic gauge field spaces. This shows that the group $K_1(C(W))$
becomes important in those situations in which the massive d=11 background
with M$9$-branes wrapped inside is subject to the dimensional reduction and
ends into a massive Type IIA background with D$8$-branes wrapped inside,
in the limit where the direction of the Killing isometry decompactifies.

If during the process of dimensional reduction the M$9$-branes reduce to
D$8$-branes, a map $\bar{\mu}$ between $Z$ and $W$ exists. If the two
reductions, of the background and of the 9-branes, respectively, hold
simultaneously and along $\hat{k}$, then we can take
$\bar{\mu} = \bar{\lambda}|$, where $\bar{\lambda}|$ is the reduction of
$\bar{\lambda}$ to $Z$. However, in our analysis we can take a more general
situation in which $\bar{\mu}$ is independent of $\bar{\lambda}|$. The
particular case emerges from this one at equality.

Let us denote by $\bar{\mu}^*_*$ the induced map between the algebraic
K-theories and consider the case $G=\tor$. Since the wrapped 8-branes are
obtained by dimensionally reducing the 9-branes, the two different
gemetrical map compositions from
$K_p (C(Z) \times_{\bar{k}^*} \tor)$
to $K_p (C(W))$ should form commutative diagrams for $p=0,1$. Using
(\ref{fintmap1}) and (\ref{fintmap}) these diagrams can be written as
\beq
\new{
\begin{array}{ccccccc}
K_0 (C(W))&  &  &  & K_1 (C(W))&  &  \\ 
{\scriptstyle{\bar{\mu}}^*_*}\uparrow  & \nwarrow {\scriptstyle{a_{*\tor}}}&
  &  &{\scriptstyle{\bar{\mu}}^*_*}\uparrow  &
   \nwarrow {\scriptstyle{b_{*\tor}}}  &  \\
K_1 (C(Z))&\buildrel{\rho_*}\over\longrightarrow&
 K_0(C(Z)\times_{\bar{k}^*}\tor )&  & K_0(C(Z)) &
  \buildrel{\rho_*}\over\longrightarrow & K_1(C(X)\times_{\hat{k}}^*\tor)
\end{array}}
\label{comdiag}
\eeq

The resulting relations between maps represent constraints on $a_{*\tor}$ and
$b_{*\tor}$ they are consequence of the condition that the D$8$-brane system
is obtained
from a M$9$-brane system. Note that these constraints do not imply
$a_{*\tor}=b_{*\tor}$ since in the two diagrams in (\ref{comdiag}) the maps
$\bar{\mu}^*_*$ and $\rho_*$ are actually splitted, acting on different
groups.

In the case $G=\zed$ we can also require the existence of the
map $\bar{\mu}$ between $Z$ and $W$ as above. The difference occurs in
that there are no physical reasons for taking $\bar{\mu} = \bar{\lambda}|$
and, more important, for assuming that the diagrams corresponding to
(\ref{comdiag}) are commutative. If $G=\real$, the above construction can
be repeated but in order to impose naturally the commutativity of the diagrams
we have to take firstly a finite radius of the Killing direction and
afterwards to let it go to infinity.

We note in the end of this section that if $X$ and $Y$ are only locally
compact, one can repeat the above analysis if we take $X^+$ and $Y^+$
instead of the original spacetime manifolds. However, if the backgrounds are
general, we must check out if the corresponding algebras of continuous
functions are commutative C$^*$-algebras with unity.

\newpage

\newsection{Discussions}

It is of interest to see if the above analysis provides any information
about all Type IIA D-brane charges which take values in the topological
K-theory group $K^{-1}(Y)$ \cite{ew,ph}. By Novodvorskii's theorem (see
Theorem 2 from Appendix) this group is isomorphic with $K_1(C(Y))$ in
which $K_1 (C(X))$ is mapped by the corresponding $\bar{\lambda}^*_*$
map. Consequentely, if $G=\tor$ we can extract from the sequence
(\ref{seqtone}) the following map
\beq
c_{*\tor} : K_{0}(C(X)\times_{{\hat{k}}^*}\tor)
\longrightarrow K_1 (C(Y)),
\label{cmap}
\eeq
where
\beq
c_{*\tor}=\bar{\lambda}^*_* \circ t_* .
\label{cmap1}
\eeq
This function maps the covariant
equivalent spaces of gauge fields from $K_{0}(C(X)\times_{{\hat{k}}^*}\tor)$,
i.e. defined on the massive d=11 supergravity background spacetime $X$ with
the toric Killing vector action, on the Type IIA D-brane charges in the
massive Type IIA background $Y$. Similar maps can be constructed for
$G=\real$ and $G=\zed$. We note now that no reference is made to any 9-brane.
As a matter of fact, the subsequence of (\ref{seqtone}) in which $c_{*\tor}$
is defined as a composition do not include any group of $Z$ as can be easily
verified. The is also true for the other two groups.

Another important problem concerns the amount of information that can be
extracted from the maps constructed in the previous section. The point here
is that in constructing (\ref{abrmaps}), (\ref{abzmaps}) and (\ref{abtmaps})
we used the Connes' Thom isomorphism, the Pimsner-Voiculescu six-term
exact sequence and its consequence (\ref{pas}). While the isomorphism
preserves the maximum of information while going from one group to another,
the same is no longer true for the cyclic six-term exact sequences since
the same group appears in two places. However, despite this limitation,
six-terms sequences are a powerfull tool in many particular cases.

To illustrate this fact, let us consider that the 9-branes are wrapped on
$S^{10}$. In this case the algebraic K-theory groups are given by
\bea
K_0 (C(S^{10})) & = & \zed / 2\zed
\nonumber\\
K_1 (C(S^{10})) & = & 0.
\label{ksphere}
\eea
By introducing (\ref{ksphere}) in the six-exact term sequence for, say,
$G=\zed$ we obtain the following relations
\bea
K_0 (C(S^{10})\times_{\bar{k}^* }\zed ) &\cong &
(\zed / 2 \zed )/ im (1- \bar{k}^*_* )
\nonumber\\
K_1 (C(S^{10})\times_{\bar{k}^*} \zed ) &\cong &
im (1- \bar{k}^*_* )
\label{ksphere1}
\eea
which contain information about the K-groups as well as the action of the
group of integers on the compact space. We can obtain similar relations
for $G=\tor$ or $G=\real$. These relations can be used to improve our
knowledge of the maps $a_*$ and $b_*$, respectively.

Another simplification of the sequences used to connect the relevant groups
appears when the manifolds on which the 8- and 9-branes
are wrapped are deformation retracts of the
corresponding spacetime manifolds.
Without presenting any details, we just note that
if we take for example
$(X,Z)$ a compact pair then $(X^+ ,Z^+ )$ is pointed since the two base
points
are identified with the point at infinity and if we further assume that
$Z^+$ is also a retract of $X^+$ the topological K-groups satisfy
\cite{mfa,mk}
\beq
\widetilde{K}^{-p}(Z^+) \cong \widetilde{K}^{-p}(X^+).
\label{retr}
\eeq
Since we have the following relations
\bea
\widetilde{K}^{-p}(X^+) &=& K_{p}(C(X))
\nonumber\\
\widetilde{K}^{-p}(Z^+) &=& K_{p}(C(Z))
\label{retr1}
\eea
supplementary relations among the maps $m_*$ and $n_*$ are given by the
following sequence

\beq
\new{
\begin{array}{ccc}
K_0 (C(Z)) &
\buildrel{m_*}\over\longrightarrow&
K_1 (C(X)) \\
\parallel  &  & \parallel  \\ 
K_1 (C(Z)) &
\buildrel{n_*}\over\longleftarrow&
K_0 (C(X))
\end{array}} 
\label{retseq}
\eeq
which connects the sequences that define $\alpha_*$ and $\beta_*$ as
composition maps.

There can be many topological configurations in which the six-term
exact sequences provide more information than in their general
formulation.
Thus we conclude that the maps between the relevant groups which were
constructed in the previous section carry signifiant information about
the system.

So far we have considered mainly noncompact Killing vectors $\hat{k}$. Let
us briefly discuss some of the particularities of the more realistic theories
with a compact Killing isometry viewed as the action of a compact group.
Consider the general case of a larger compact group of isometries $G$ of
both $X$ and $Z$ which are also compact manifolds
\footnote{In the case when $K=\hat{k}$ the group is $G=U(1)$.}.

Let us denote by $K$ and action of $G$ on $X$ and by $\bar{K}$ an action on
$Z$. To all actions $\{ K \}$ correspond actions of $G$ on $C(X)$ and on the
vector bundles $E\rightarrow X$. A $G$-vector bundle is a vector bundle $E$
for which the $G$-action is induced in a way that the projection
$E\rightarrow X$ is equivariant. Like for usual vector bundles one can
establish an exact correspondence between $G$-vector bundles and finitely
generated projective modules of $C(X)\times_{K^*}G$. In particular, there are
$G$ K-theory groups the following isomorphisms hold \cite{bb}
\bea
\gamma^G_* &:& K^G_0 (C(X)) \longrightarrow K_0 (C(X)\times_{K^*}G)
\nonumber\\
d^G_* &:& K^G_0 (C(X)) \longrightarrow K_0 (C(X)),
\label{kgth}
\eea
where $K^G_0 (X)$ is the abelian group of equivalence classes of $G$-vector
bundles and $K^G_0 (C(X))$ is the abelian (Grothendieck) group of
equivalence classes of finitely generated $G$-projective modules over
$C(X)$. The same construction can be done for $Z$ and we denote the
isomorphisms (\ref{kgth}) in this case with $\chi^G_*$ and $e^G_*$,
respectively.

In analogy with the case studied in the previous section, we would like
to find a map between the $G$-covariant gauge fields on M$9$-branes
and the D$8$-brane charges. Proceeding along the same line, we
denote by $i^G_*$ the map induced by the natural inclusion of $Z$ in $X$
in K-theory. It is not difficult to see that by using $\bar{\lambda}^*_*$
discused in the previous section we obtain the following sequence
\beq
K_0 (C(Z)\times_{\bar{K}^*}G)~\longrightarrow ~ K^G_0 (C(Z))~
\longrightarrow ~ \cdots~ \longrightarrow ~ K_0 (C(Y))~
~\longrightarrow ~ K_1 (C(W)).
\label{lastseq}
\eeq
Unfortunately, this sequence defines a map between the equivalence classes 
of spaces of $G$-covariant gauge fields on M$9$-branes from
$K_0 (C(Z)\times_{\bar{K}^*}G)$ and the group $K_1 (C(W))$. The result is
unwanted since we do not know how to interpret the later group in terms of
8-branes.

In order to solve this problem we pick up a subalgebra $A$ of $C(W)$ which
makes the following split sequence exact
\beq
0~ \longrightarrow ~ A~\buildrel{\iota^A}\over\longrightarrow ~
\buildrel{\pi^A}\over\longrightarrow ~ C(X)/A ~
\longrightarrow ~0
\label{splseq}
\eeq
but otherwise arbitrary. The standard exact sequence theorem \cite{es,bb}
states that the associated six-term cyclic sequence is exact. From this we
can extract the K-theory map
\beq
\delta^A_* = \iota^A_* \circ \bar{\partial} \circ \pi^A_* ,
\label{deltaA}
\eeq
where $\bar{\partial}$ is the composition of the suspended index map
$\partial : K_2 (C(X)/A) \rightarrow K_1 (A)$ with the Bott map. Note that 
$\delta^A_*$ is independent of $G$ but depends on the choice of $A$.
Using $\delta^A_*$ we can map the equivalence classes of spaces of $G$-vector
fields on the M$9$-branes on the D$8$-brane charges. Explicitely,
we have the following map
\beq
\Delta^{GA}_* : K_0 (C(Z)\times_{\bar{K}^*}G) \longrightarrow
K_0 (C(W))
\label{bdfin}
\eeq
where the expresion of $\Delta^{GA}_*$ in terms of components is given by
\beq
\Delta^{GA}_* = (\delta^A_*)^{-1} \circ v_*^{-1} \circ
\bar{\lambda}^*_* \circ d^G_* \circ i^G_* \circ (e^G_*)^{-1}
\circ (\chi^G_*)^{-1} .
\label{bdfin1}
\eeq
We note that, in general, the map (\ref{bdfin1}) cannot be completely
satisfactory because it depends on the arbitrary algebra $A$. From the
physical point of view we can speculate that $A$ should be identified with
the algebra of continuous functions on some compact submanifold
$ U \subset W$ on which some lower D$p$-branes are wrapped on.
In any case, the choice of a submanifold $U$ reduces further the symmetries
of the system.

\bigskip

\noindent
{\bf Acknowledgements:} We would like to thank to S. Sorella,
J. A. Helayel-Neto, F. Toppan, B. Schroer for discussions and
R. Exel for correspondence. It is a pleasure to thank LNCC and
UCP for hospitality while the manuscript was being completed.
I also acknowledge a FAPERJ postdoc fellowship.
\newpage

\appendix{\bf Appendix}

In this Appendix we collect two classical mathematical results. The first
theorem is a consequence of Bott's Theorem for topological K-theory of
pointed compact pairs.
\begin{theorem}
Let $(M,N)$ be a pointed compact pair and let $i:N\rightarrow M$,
$j:(M, \emptyset)\rightarrow (M,N)$ denote the natural inclusions. In 
the following diagram

\beq
\new{
\begin{array}{llllllllll}
\cdots \rightarrow & K^{-2}(M) & \buildrel{i^*}\over\rightarrow &
K^{-2}(N)  & \buildrel{\delta}\over\longrightarrow & K^{-1}(M,N)
&  \buildrel{j^*}\over\rightarrow & K^{-1}(M) &  &  \\
&  &  &  {\scriptstyle\beta}\nwarrow
&  & {\scriptstyle\delta \circ \beta}\nearrow  &
&  & {\scriptstyle{i^*}}\searrow  & \\
&  &  &  & K^{0}(N)&  &  &  &  & K^{1}(N) \\ 
&  &  &  &  & {\scriptstyle{i^*}}\nwarrow  &
&  & {\scriptstyle\delta}\swarrow  &  \\
&  &  &  &  &  K^0 (M) & \buildrel{j^*}\over\leftarrow & K^0 (M,N) &  &  
\end{array}}, 
%\label{bottc}
\eeq
where the top row is the exact sequence of the pair $(M,N)$. Then the
hexagonal
part of the diagram is exact. Here, $\delta$ is the coboundary map
\beq
\delta : K^{-p}(N) \longrightarrow K^{-p+1}(M,N)
%\label{cobound}
\eeq
and $\beta$ is the Bott isomorphism
\beq
\beta : K^{-p}(M,N) \longrightarrow K^{-(p+2)}(M,N).
%\label{bottis})
\eeq
\end{theorem}

The second theorem is a particular case of a theorem by Novodvorskii
\cite{men,jlt} which establishes under what circumstances the topological and
algebraical K-theory groups of a commutative Banach algebra are equivalent.
In our case, the algebras of continuous functions in the theorem below are
C$^*$-algebras which implies that they are also Banach algebras.
\begin{theorem}
The Gel'fand transform $C_0 (M) \rightarrow C(\widehat{(C_0 (M)})$, where
$M$ is a locally compact space, induces an isomorphism
\beq
\widetilde{K}_p (C_0 (M)) \cong \widetilde{K}^{-p}(M^+ )~~,~~p=0,1.
\label{novod1}
\eeq
Similarly, the Gel'fand transform $C(M) \rightarrow C( \widehat{C(M)})$, where
$M$ is a compact space, induces an isomorphism
\beq
K_p (C_0 (M)) \cong K^{-p}(M)~~,~~p=0,1.
\label{novod2}
\eeq
\end{theorem}
Here, $\widehat{C(M)}$ is the Gel'fand space of the algebra $C(M)$, i. e.
the space of equivalence classes of irreducible representations of $C(M)$
(or the space of its maximal ideals). This space can be identified setwise
as well as topologically with $M$. For any element $f\in C(M)$ its Gel'fand
transform is the complex valued function $\hat{f} : \widehat{C(M)}
\rightarrow \complex $ given by $\hat{f}(\varphi ) = \varphi (f)$ for
any $\varphi \in \widehat{C(M)}$.

\end{document}